\documentclass[aps,prl,preprint,groupedaddress]{revtex4-1}

\usepackage{graphicx}
\usepackage{dcolumn}
\usepackage{bm}
\usepackage{color}

\begin{document}


\title{Experimental Realization of a Quantum Breathing Pyrochlore Antiferromagnet}

\author{
K. Kimura$^1$, S. Nakatsuji$^{2,3}$, and T. Kimura$^{1}$
}
\affiliation{
$^1$Division of Materials Physics, Graduate School of Engineering Science, Osaka University, Toyonaka, Osaka 560-8531, Japan\\
$^2$Institute for Solid State Physics (ISSP), University of Tokyo, Kashiwa, Chiba 277-8581, Japan\\
$^3$PRESTO, Japan Science and Technology Agency (JST), 4-1-8 Honcho Kawaguchi, Saitama 332-0012, Japan
}

\date{\today}

\begin{abstract}
The synthesis and characterization of a new Yb-based material Ba$_3$Yb$_2$Zn$_5$O$_{11}$ are reported. This material is identified as the first model system of a pseudospin-1/2 quantum antiferromagnet on ``breathing'' pyrochlore lattice characterized by an alternating array of small and large Yb tetrahedra. Despite dominant antiferromagnetic interactions $J\sim7$ K, a large amount of magnetic entropy (25\%) remains at 0.38 K, indicating that each small Yb tetrahedron forms a unique doubly degenerate singlet state. These results are well described by the Heisenberg pseudospin-1/2 single tetrahedron model.

\begin{description}
\item[PACS numbers]
75.40.Cx, 75.10.Kt, 75.50.-y
\end{description}
\end{abstract}


\maketitle


Search for novel and exotic phenomena associated with spin degrees of freedom has been central to condensed matter physics \cite{Lee2008, Balents2010}. One of the most attractive systems in three dimension is a pyrochlore lattice magnet, which consists of corner-sharing regular tetrahedra of magnetic ions \cite{Subramanian1983}. The inherent geometrical frustration suppressing a conventional magnetic order often leads to various unusual properties \cite{GardnerReview}. Experimental examples include the spin-driven lattice distortion in chromium spinels $M$Cr$_{2}$O$_{4}$ \cite{Lee2000,Chung2005} and spin ice (-like) state in rare-earth pyrochlore oxides $Re_{2}B_{2}$O$_{7}$ \cite{Harris1997,Ramirez1999,Machida2010,Ross2011,Kimura2013}.

Spin-1/2 quantum pyrochlore Heisenberg antiferromagnets are known to be promising candidates for three-dimensional quantum spin liquid \cite{Zheng2005,Canals1998,Tsunetsugu2001,Tsunetsugu2001b,Berg2003,Moessner2006,JH-Kim2008,Burnell2009}. Despite numerous experimental and theoretical efforts, their ground state properties have not yet been established because of the lack of a model material and the unavailability of exact solutions.
A popular theoretical approach to this problem is to first decouple the full ($i.e.$, uniform) pyrochlore lattice into a set of independent tetrahedra and then reconnect them perturbatively. Based on this approach, several types of singlet ground states were proposed \cite{Canals1998,Tsunetsugu2001,Tsunetsugu2001b,Berg2003}. For example, Tsunetsugu found a non-chiral dimerized ground state with a four sublattice structure \cite{Tsunetsugu2001b}. However, a question arises whether this cluster approach correctly describes the true ground state. Indeed, subsequent studies based on the fermionic mean field theory have suggested a chiral spin liquid state \cite{JH-Kim2008,Burnell2009}.
Therefore, a material composed of a regular spin-tetrahedral unit with an inter-tetrahedron coupling is of great interest because understanding these inter-tetrahedron couplings is expected to provide important insights on this issue. Moreover, such a material may show exotic magnetism based on the unique properties of the single tetradedron associated with the spin chirality.

Recently, appropriate materials have been found in the spinel family, Li$A'$Cr$_4$O$_8$ ($A'= $ In, Ga), which consist of an alternating array of small and large Cr$^{3+}$ tetrahedra (spin-3/2); hence they are named ``breathing'' pyrochlore lattice \cite{Okamoto2013}. The tunable ratio of exchange interactions in large ($J'$) and small ($J$) tetrahedra, $J'/J$, makes these system suitable for studying inter-tetrahedral coupling effects \cite{Okamoto2013}. However, no breathing pyrochlore antiferromagnet with quantum spin-1/2 has been reported to date.

\begin{figure}[t]
\includegraphics[width=10cm]{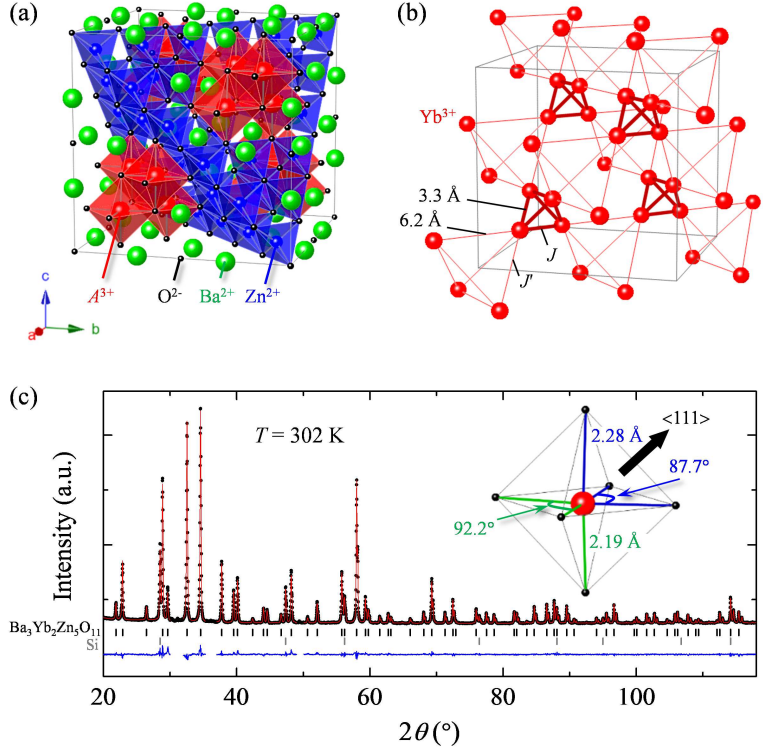}
\caption{(color online).
(a) Crystal structure of Ba$_3$$A$$_2$Zn$_5$O$_{11}$ ($A$ = Lu and Yb). $A$$_4$O$_{16}$ cluster and Zn$_{10}$O$_{20}$ super-tetrahedron are depicted. (b) Breathing pyrochlore lattice formed by $A^{3+}$ ions. Inter- and intra-tetrahedron distances are given for $A$ = Yb$^{3+}$. (c) X-ray diffraction pattern (black circles) of Ba$_3$Yb$_2$Zn$_5$O$_{11}$ in the presence of a silicon standard at room temperature. The red line corresponds to the best fit from the Rietveld refinement based on the structural model of the Lu analogue \cite{Rabbow1996}. Upper and lower vertical marks denote the Bragg peak positions for Ba$_3$Yb$_2$Zn$_5$O$_{11}$ and the silicon standard, respectively. The bottom line represents the difference between experimental and calculated intensities. Inset: Local environment of Yb$^{3+}$ and associated geometrical parameters. The crystalline electric field formed by six surrounding O$^{2-}$ ions exhibits a cubic-like symmetry with a small trigonal distortion along the $\langle$111$\rangle$ direction.
\label{Struct}}
\end{figure}

In this Rapid Communication, we show the new material Ba$_3$Yb$_2$Zn$_5$O$_{11}$ to be a model system of a quantum breathing pyrochlore lattice antiferromagnet. It belongs to a family of Ba$_3$$A$$_2$Zn$_5$O$_{11}$-type compounds ($A$ = trivalent ion). The crystal structure of this family was solved by the single crystal X-ray diffraction (XRD) technique \cite{Scheikowski1993,Rabbow1996}. They crystallize in the unique structure with the cubic space group $F\bar{4}$3$m$, in which the $A$$_4$O$_{16}$ cluster and Zn$_{10}$O$_{20}$ super-tetrahedron align alternatively and Ba ions fill the interstices, as depicted in Fig.~\ref{Struct}(a). Interestingly, $A$ sites form a breathing pyrochlore lattice [Fig.~\ref{Struct}(b)], although all reported compounds of this family have $non$-$magnetic$ $A$ ions ($A$ = In, Lu) \cite{Scheikowski1993,Rabbow1996}. Here we successfully synthesized Ba$_3$Yb$_2$Zn$_5$O$_{11}$, where the breathing pyrochlore lattice is formed by Yb$^{3+}$ ions with $magnetic$ Kramers doublets carrying pseudospin-1/2. Our magnetic and thermodynamic measurements revealed the formation of a unique spin-singlet state with a double degeneracy that can be labelled by scalar spin chirality.

Polycystalline samples of Ba$_3$Yb$_2$Zn$_5$O$_{11}$ were prepared by the standard solid state reaction method. A stoichiometric mixture of BaCO$_3$, Yb$_2$O$_3$, and ZnO was heated at 1150　$^\circ$C for 100 hours with several intermediate grindings. The powder XRD pattern of the sample in the presence of a silicon standard (NIST 640d) was recorded by a RINT-2100 diffractometer (Rigaku) with Cu K$\alpha$ radiation at room temperature ($T$) and 20 K. Magnetization measurements down to 1.8 K and up to 7 T were performed using a commercial superconducting quantum interference device (SQUID) magnetometer (Quantum Design, MPMS).
The specific heat was measured down to 0.38 K by means of a thermal relaxation method using a commercial calorimeter (Quantum Design, PPMS). 

Figure \ref{Struct}(c) shows the XRD pattern taken at room temperature. 
The data was analyzed by the Rietveld method using the PDXL software (Rigaku). The crystal structure of Ba$_3$Lu$_2$Zn$_5$O$_{11}$ \cite{Rabbow1996} was used as a starting model and, subsequently, the atomic positions and isotropic atomic displacement parameters of Ba, Yb, and Zn sites were refined.
Three $2\theta$ ranges exhibiting very weak peaks for impurities were excluded from the refinement \cite{BYbZOimpurity2}. The good agreement between observed and calculated patterns ($R_{\rm wp} = 5.40$, $R_{\rm p} = 4.21$, $S = 1.52$) ensures the Ba$_3$$A$$_2$Zn$_5$O$_{11}$-type structure realized in the present compound, in which Yb$^{3+}$ ions form a breathing pyrochlore lattice. 
The lattice constant 13.4871(1) {\AA}  is larger than the value for the Lu analogue 13.452 \AA~\cite{Rabbow1996}, in agreement with the size of ionic radius Yb$^{3+}$$>$Lu$^{3+}$. Intra- and inter- Yb-tetrahedron distances are determined to be $\sim$3.29 {\AA} and $\sim$6.23 {\AA} [Fig.~\ref{Struct}(b)].
The XRD pattern at 20 K did not show any peak splitting, confirming the absence of structural phase transition at least down to this temperature. 

\begin{figure}[t]
\includegraphics[width=10cm]{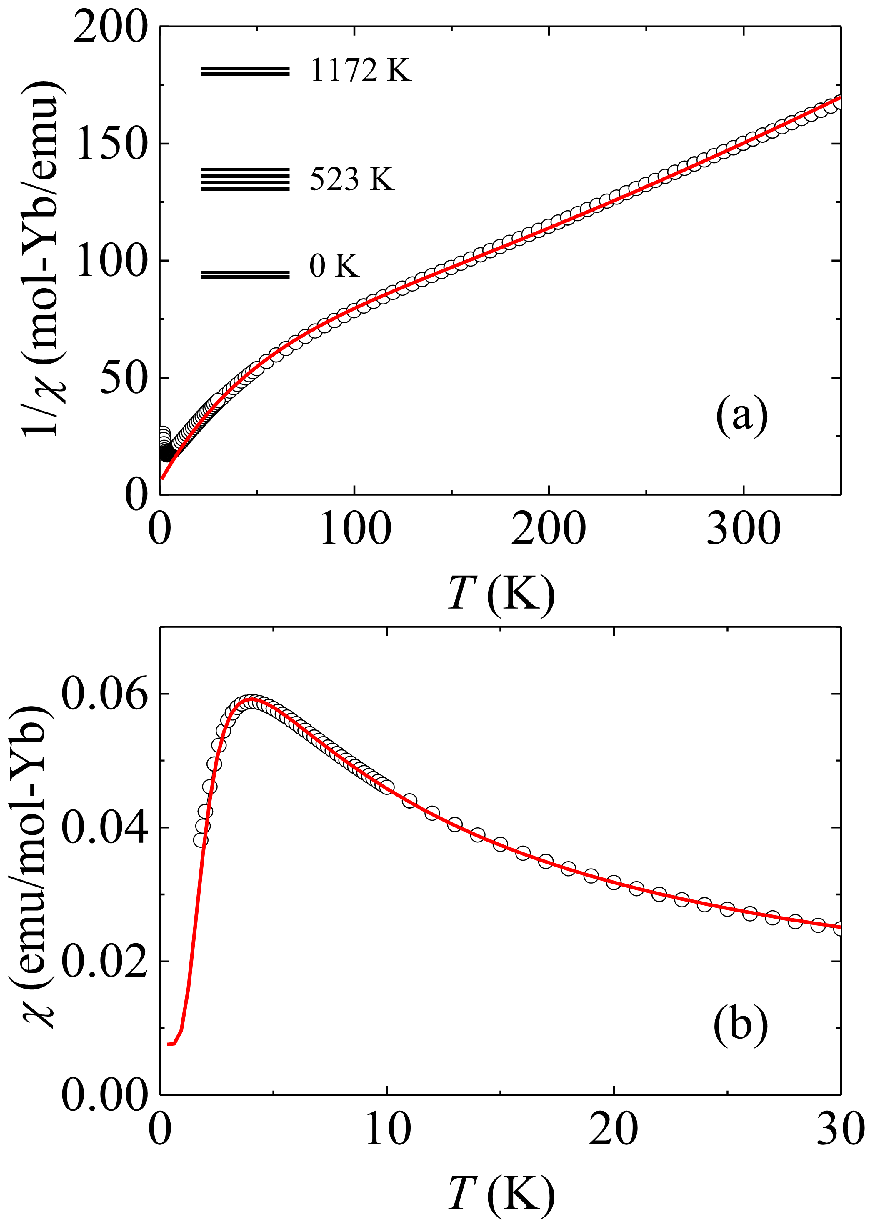}
\caption{(color online).
(a) Temperature dependence of the inverse magnetic susceptibility $1/\chi(T)$ measured at a field of 0.1 T for $T<30$ K and 1 T for $T>30$ K. The red line is the calculated curve based on the cubic crystalline electric field (CEF). The corresponding CEF scheme is illustrated. (b) $\chi(T)$ below $T=30$ K. The red line shows the fit of the single tetrahedral Heisenberg model (see text). 
\label{chi}}
\end{figure}

Analyses of the crystalline electric field (CEF) scheme of  Yb$^{3+}$ ions (4$f^{13}$) and magnetic susceptibility data provide strong evidence for the realization of a quantum breathing pyrochlore antiferromagnet. The true site symmetry of the Yb$^{3+}$ ions (3$m$) requires six independent CEF parameters in the effective CEF Hamiltonian. However, a close look at the local environment [inset, Fig.~\ref{Struct}(c)] shows small differences between Yb-O bond lengths [2.28 $\rm \AA$ (blue) vs. 2.19 $\rm \AA$ (green)] and O-Yb-O angles [87.6$^\circ$ (blue) vs. 92.2$^\circ$ (green)], indicating that the trigonal distortion along the $\langle$111$\rangle$ direction from the cubic octehedral O$^{2-}$ coordination is relatively small.
The CEF can thus be approximated by the cubic octahedral symmetry, and the resultant effective CEF Hamiltonian is written as \cite{Hutchings1964}
\begin{eqnarray}
H_{\rm CEF}= &&(-2/3)B_4[O_4^0-20\sqrt{2}O_4^3]+(16/9)B_6[O_6^0+35\sqrt{2}/4O_6^3+(77/8)O_6^6],
\label{eqHcef}
\end{eqnarray}
where the trigonal axis is taken as the quantized axis, and $B_n$ are the CEF parameters and $O_n^m$ are the Stevens operator equivalents \cite{Stevens1952}. The point charge model gives $B_4<0$ and $B_6>0$ for Yb$^{3+}$ in the octahedral coordination \cite{Lea1962}. To verify this approximation, the calculated magnetic susceptibility $\chi$ was compared with experimental results using the form of $\chi=\chi_{\rm dia}+\chi_{\rm CEF}/(1+{\lambda}\chi_{\rm CEF})$, where $\chi_{\rm dia}$ is the core diamagnetic susceptibility fixed to be $-4.13\times10^{-4}$ emu/mol-Yb \cite{VanVleck1932}, $\lambda$ is a parameter describing uniform exchange interactions, and $\chi_{\rm CEF}$ is the single ion CEF susceptibility given by the Van-Vleck formula \cite{VanVleck1932}:
\begin{eqnarray}
  \chi_{\rm CEF}=
 \frac{N_{\rm A}g_{J}^2 \mu_{\rm B}^2}{k_{\rm B} \sum_{n} e^{-E_{n}/T}}
 \biggl(
 \frac{\sum_{n} {|\langle n|\bf{J} \it{|n} \rangle|}^2 e^{-E_{n}/T}}{T}
+\sum_{n}\sum_{m \ne n}{|\langle m|\bf{J} \it{|n} \rangle|}^2 \frac {e^{-E_{n}/T}-e^{-E_{m}/T}}{E_{m}-E_{n}}
\biggr).
\label{eqchicef}
\end{eqnarray}
In Eq.~(\ref{eqchicef}), $N_{\rm A}$ is the Avogadro's number, $k_{\rm B}$ is the Boltzmann constant, $\mu_{\rm B}$ is the Bohr magneton, $g_{J}$ is the Lande's $g$-factor ($g_{J} = 8/7$), $\bf{J}$ is the angular momentum operator, and $E_{n}$ is the energy of the $n$th level in units of $T$.
The red line in Fig.~\ref{chi}(a) represents the best fit for $T>20$ K with $B_4=-0.6$ K, $B_6=0.002$ K, and $\lambda=-5.0$ emu/mol (antiferromagnetic) and shows a good agreement between experiment and calculations, validating the cubic approximation. The obtained CEF scheme is illustrated in Fig.~\ref{chi}(a). The ground state corresponds to a magnetic Kramers doublet with an effective $g$-factor ($g_{\rm eff}$) of 2.66 for pseudospin-1/2, which is fully-isotropic because of the cubic approximation. The very large gap ($>500$ K) between ground state and excited states (a quartet and a doublet) ensures that the low-temperature properties are described by pseudospin-1/2.

To examine correlations among pseudospin-1/2, we turn to $\chi(T)$ at low temperature. For $T<30$ K, $\chi(T)$ shows no signs of conventional long-range ordering down to 1.8 K but exhibits a broad maximum at around 4 K [Fig.~\ref{chi}(b)]. For $10$ K $<T<$ 30 K, the data obeys the Curie-Weiss law, $\chi(T)=C/(T-\theta_{\rm CW})+\chi_{0}$, where the constant term $\chi_{0}$ is the sum of $\chi_{\rm dia}$ and the Van-Vleck contribution $\chi_{vv}=7.3\times10^{-3}$ emu/mol-Yb calculated from the CEF scheme \cite{VanVleck1932}. The fit yields a negative Weiss-temperature $\theta_{\rm CW}=-6.7(1)$ K, indicative of sizable antiferromagnetic interactions among pseudospin-1/2. The $g_{\rm eff}$ value of 2.66 calculated from the Curie constant $C$ is consistent with the CEF analysis. 

\begin{figure}[t]
\includegraphics[width=10cm]{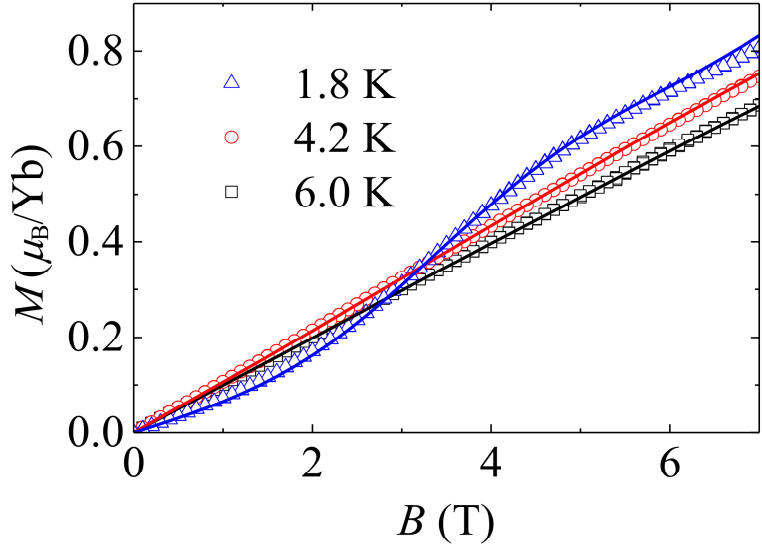}
\caption{(color online).
Magnetic field dependence of the magnetization at selected temperatures. Open black squares (6.0 K), red circles (4.2 K), and blue triangles (1.8 K) represent experimental data. Solid lines correspond to values calculated using Eq.~(\ref{eqHtetra}) with a set of parameters obtained from the fitting of $\chi(T)$.
\label{MH}}
\end{figure}

These results therefore establish that Ba$_3$Yb$_2$Zn$_5$O$_{11}$ is a quantum psudospin-1/2 breathing pyrochlore antiferromagnet. The next task is to characterize the magnetism of this unique spin system in more detail. Because the Yb-Yb distance differs significantly in intra- and inter-tetrahedra [Fig.~\ref{Struct}(c)], intra-tetrahedon couplings $J$ is expected to largely surpass inter-tetrahedron couplings $J'$. Therefore, the broad maximum in $\chi(T)$ observed at around 4 K [Fig.~\ref{chi}(b)] suggests the formation of a quantum spin singlet state in a small Yb-tetrahedron. This is supported by the magnetization curves at selected temperatures (Fig.~\ref{MH}). Although linear above 4 K, the magnetization curve shows a clear non-linear increase at $B\sim3$ T below 4 K---a signature of the singlet-triplet crossover.

In the single tetrahedron approximation, the effective Hamiltonian for the pseudospin-1/2 is written as
\begin{eqnarray}
\mathcal{H}_{\rm tetra}= -J\sum^{}_{i<j} \bf{S}_{\it i} \cdot \bf{S}_{\it j}+\it g_{\rm eff}\mu_{\rm B}\bf{H} \it \cdot \sum^{}_{i}\bf{S}_{\it i},
\label{eqHtetra}
\end{eqnarray}
where $\bf{S}_{\it i}$ is the pseudospin-1/2 operator of the ${i}$th Yb$^{3+}$ ion in a small tetrahedron ($i=$ 1, 2, 3, and 4), $J$ is the Heisenberg exchange interactions, and $\bf H$ is an external magnetic field.
At $\bf H \rm = 0$, the quantum states of the tetrahedron are characterized by the total pseudospin, $S_{\rm T}$ = 0, 1, or 2, where $S_{\rm T} \equiv \sum^{}_{i}S_{i}$. For antiferromagnetic $J<0$, the ground state is a $doubly$ degenerate singlet with $S_{\rm T}=0$ and energy $\epsilon =3J/2$, the first excited state a triply degenerate triplet with $S_{\rm T}=1$ and $\epsilon =J/2$, and the next excited state a nondegenerate quintet with $S_{\rm T}=2$ and $\epsilon =-3J/2$.

To compare the experimental results and model, the susceptibility is defined as $\chi(T) =  \chi_{\rm tetra}(T)+\chi_{0}$, where $\chi_{0}$ is the constant term and $\chi_{\rm tetra}(T)$ is the intrinsic susceptibility of the single tetrahedron. Given the energy levels as above, $\chi_{\rm tetra}(T)$ per Yb ion is expressed as
\begin{eqnarray}
  \chi_{\rm tetra}(T) =
  \frac{N_{\rm A}g_{\rm eff}^2 \mu_{\rm B}^2}{2k_{\rm B}T}
  \frac{5+3e^{-2J/T}}{5+9e^{-2J/T}+2e^{-3J/T}}.
\label{eqchitetra}
\end{eqnarray}
In the fitting procedure, we treat $J$, $g_{\rm eff}$, and $\chi_{0}$ as adjustable parameters. A good agreement is obtained between the model and experiment for $T<30$ K with $J=-6.43(1)$ K, $g_{\rm eff}=2.569(3)$, and $\chi_{0}=7.5(1)\times10^{-3}$ emu/mol-Yb, as indicated by the red line in Fig.~\ref{chi}(b).  This $g_{\rm eff}$ value is consistent with the CEF calculation and $\chi_{0}$ is comparable with the sum of $\chi_{\rm dia}$ and $\chi_{vv}$. Moreover, the magnetization curves above 1.8 K are successfully reproduced with the same set of parameters obtained from $\chi(T)$ (solid lines, Fig.~\ref{MH}).

\begin{figure}[t]
\includegraphics[width=10cm]{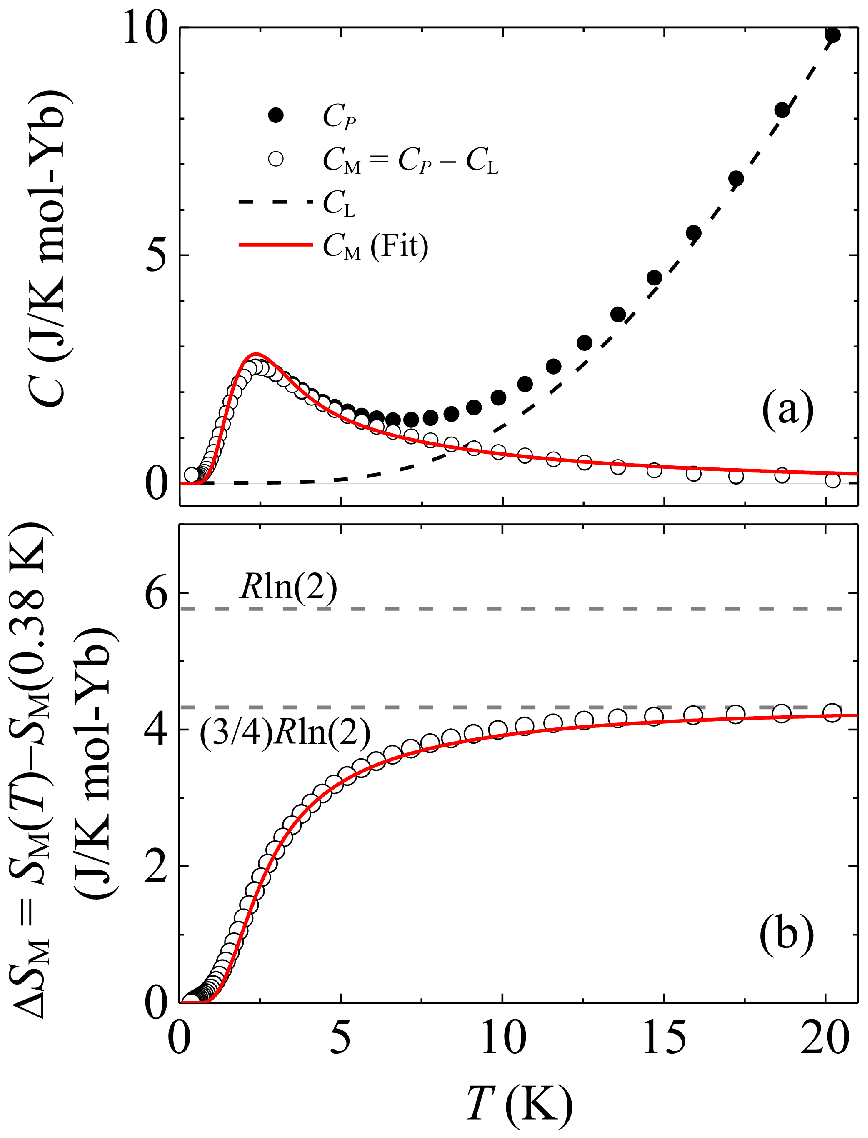}
\caption{(color online).
(a) Temperature dependence of the magnetic specific heat $C_{\rm M}$ (open circles) after subtracting the lattice contribution $C_{\rm L}$ (dashed line) from the measured specific heat $C_{P}$ (closed circles). The red solid line is a fit of Eq.~(\ref{eqCmtetra}), which yields $J=-7.1$ K. (b) Corresponding magnetic entropy expressed as a variation from $T = 0.38$ K, $\Delta S_{\rm M} = S_{\rm M}(T)-S_{\rm M}(0.38\, \rm K)$. The solid line is calculated using Eq.~(\ref{eqCmtetra}) with $J=-7.1$ K. Dashed lines represent the entropy for a two-level system $R {\rm ln}(2)$ and a single tetrahedron model $(3/4)R {\rm ln}(2)$. The difference between the two corresponds to the remaining entropy due to the singlet degeneracy.
\label{Cp}}
\end{figure}

Further information on low-$T$ properties is provided by the magnetic specific heat $C_{\rm M}$ down to $T=0.38$ K after subtracting the lattice contribution $C_{\rm L}$ from the measured specific heat $C_{P}$ [Fig.~\ref{Cp}(a)]. $C_{\rm L}$ was approximated by the specific heat of the non-magnetic analogue Ba$_3$Lu$_2$Zn$_5$O$_{11}$. $C_{\rm M}$ exhibits a broad peak associated with the singlet formation without any signs of long-range order. The corresponding magnetic entropy $S_{\rm M}$ obtained by integrating $C_{\rm M}(T)/T$ from 0.38 K to 20 K is shown in Fig.~\ref{Cp}(b). The saturated value at 20 K is close to 75\% of the value expected for a standard two level system $R\rm {ln}(2)$, and 25\% of magnetic entropy remains below 0.38 K. This value is fully consistent with the double degeneracy of the singlet ground state expected for the single tetrahedron model [Eq.~(\ref{eqHtetra})].
In this model, $C_{\rm M}$ is given by
\begin{eqnarray}
  C_{\rm M} =
  \frac{9N_{\rm A}k_{\rm B} J^2}{2T^2}
  \frac{10e^{-2J/T}+5e^{-3J/T}+e^{-5J/T}}{(5+9e^{-2J/T}+2e^{-3J/T})^2}.
  \label{eqCmtetra}
\end{eqnarray}
Experimental $C_{\rm M}$ is well reproduced by Eq.~(\ref{eqCmtetra}) by using only one adjustable parameter $J$ [red line, Fig.~\ref{Cp}(a)]. The fit yields $J=-7.1(1)$ K, comparable to $J=-6.43(1)$ K obtained from $\chi(T)$.

All the data presented here clearly demonstrate that the Heisenberg pseudospin-1/2 single tetrahedron model correctly approximates the low-$T$ properties of Ba$_3$Yb$_2$Zn$_5$O$_{11}$ \cite{BYbZOmultipole2}\nocite{Thompson2011,Hayre2013}.
One characteristic feature of this model is the double degeneracy of the singlet ground state. This ground state is interesting because it can be written by the complex chiral basis states, which are actually eigenfunctions of the scalar spin chirality defined as $\bf{S_{\it i}}\cdot (\bf{S_{\it j}}\times \bf{S_{\it k}})$ \cite{Tsunetsugu2001,Tsunetsugu2001b}. This suggests that Ba$_3$Yb$_2$Zn$_5$O$_{11}$ may possess the degree of freedom associated with the spin chirality at least down to $T=0.38$ K.

An intriguing question is the lifting process of the singlet degeneracy. Small extra exchange interactions that are symmetrically allowed in a single tetrahedron with  $T_d$ symmetry \cite{Curnoe2008,McClarty2009} do not lift any singlet degeneracy because they are transformed as a basis of the $E$ representation of the $T_{d}$ group \cite{Tsunetsugu2001b}.
One possible scenario is spontaneous lattice distortion resulting from magneto-elastic coupling. This so-called spin Jahn-Teller mechanism was originally proposed to explain the structural transition in frustrated spin-1 vanadium spinels \cite{Yamashita2000} and, subsequently, in spin-3/2 chromium spinels \cite{Tchernyshyov2002}. Here, a cubic to tetragonal structural phase transition is expected \cite{Yamashita2000}, resulting in a three-dimensional valence bond crystal. A more interesting scenario is based on inter-tetrahedron interactions, $J'$. A theoretical study on the Heisenberg $J-J'$ model has proposed a non-chiral spin-singlet order with novel singlet excitations \cite{Tsunetsugu2001b}. Alternatively, given the finite anisotropy in the real material, other types of exotic states such as chiral spin liquid may be expected. Although $J'$ is supposed to be much smaller than $J$ because of the large distance between neighboring tetrahedra $\sim 6$ {\AA} [Fig.~\ref{Struct}(b)], it is in principle non-negligible. At present, the ground state of Ba$_3$Yb$_2$Zn$_5$O$_{11}$ cannot be determined. However, this compound is expected to illustrate novel physics which has never been precedented experimentally. Specific heat, magnetization, and neutron scattering measurements below $T=0.4$ K are highly desired to examine its ground state.

Furthermore, two important aspects of Ba$_3$Yb$_2$Zn$_5$O$_{11}$ deserve attention. First, because Ba$_3$Yb$_2$Zn$_5$O$_{11}$ presents a unique breathing pyrochlore structure, understanding the effects of the inter-tetrahedral coupling $J'$ promises to provide information on the physics of uniform pyrochlore lattice antiferromagnets. This Yb-based compound benefits from an appropriate strength of the dominant exchange energy $J \sim 7$ K, facilitating access to a full phase diagram by a laboratory magnetic field. Second, Ba$_3$Yb$_2$Zn$_5$O$_{11}$ is a rare example that consists of the regular spin-tetrahedra, except for the famous spinel and pyrochlore compounds. Therefore, the present study stimulates further exploration of novel frustrated spin systems.

In summary, through the analyses of the crystalline electric field scheme and the magnetic susceptibility data, we have shown that the pseudospin-1/2 antiferromagnetic breathing pyrochlore lattice is realized in the new compound Ba$_3$Yb$_2$Zn$_5$O$_{11}$. Low-temperature thermodynamic measurements revealed that each small Yb tetrahedron forms unique doubly degenerate singlet state at $T=0.38$ K. The lifting mechanism of the degeneracy appears to involve new physics, which may relate to the exotic states of the quantum pyrochlore antiferromagnet.

\begin{acknowledgments}
We thank H. Kawamura, T. Sakakibara, T. Shimokawa, H. Tsunetsugu, and Y. Wakabayashi for helpful discussions. We also thank H. Tada for specific heat measurements. This work was partially supported by Grants-in-Aid for Scientific Research (No. 25707030) and by PRESTO of JST, Japan. Part of this work was carried out under the Commission Researcher's Program of
the Institute for Solid State Physics, the University of Tokyo.
\end{acknowledgments}

%


\begin{thebibliography}{34}%
\makeatletter
\providecommand \@ifxundefined [1]{%
 \@ifx{#1\undefined}
}%
\providecommand \@ifnum [1]{%
 \ifnum #1\expandafter \@firstoftwo
 \else \expandafter \@secondoftwo
 \fi
}%
\providecommand \@ifx [1]{%
 \ifx #1\expandafter \@firstoftwo
 \else \expandafter \@secondoftwo
 \fi
}%
\providecommand \natexlab [1]{#1}%
\providecommand \enquote  [1]{``#1''}%
\providecommand \bibnamefont  [1]{#1}%
\providecommand \bibfnamefont [1]{#1}%
\providecommand \citenamefont [1]{#1}%
\providecommand \href@noop [0]{\@secondoftwo}%
\providecommand \href [0]{\begingroup \@sanitize@url \@href}%
\providecommand \@href[1]{\@@startlink{#1}\@@href}%
\providecommand \@@href[1]{\endgroup#1\@@endlink}%
\providecommand \@sanitize@url [0]{\catcode `\\12\catcode `\$12\catcode
  `\&12\catcode `\#12\catcode `\^12\catcode `\_12\catcode `\%12\relax}%
\providecommand \@@startlink[1]{}%
\providecommand \@@endlink[0]{}%
\providecommand \url  [0]{\begingroup\@sanitize@url \@url }%
\providecommand \@url [1]{\endgroup\@href {#1}{\urlprefix }}%
\providecommand \urlprefix  [0]{URL }%
\providecommand \Eprint [0]{\href }%
\providecommand \doibase [0]{http://dx.doi.org/}%
\providecommand \selectlanguage [0]{\@gobble}%
\providecommand \bibinfo  [0]{\@secondoftwo}%
\providecommand \bibfield  [0]{\@secondoftwo}%
\providecommand \translation [1]{[#1]}%
\providecommand \BibitemOpen [0]{}%
\providecommand \bibitemStop [0]{}%
\providecommand \bibitemNoStop [0]{.\EOS\space}%
\providecommand \EOS [0]{\spacefactor3000\relax}%
\providecommand \BibitemShut  [1]{\csname bibitem#1\endcsname}%
\let\auto@bib@innerbib\@empty
\bibitem [{\citenamefont {Lee}(2008)}]{Lee2008}%
  \BibitemOpen
  \bibfield  {author} {\bibinfo {author} {\bibfnamefont {P.~A.}\ \bibnamefont
  {Lee}},\ }\href@noop {} {\bibfield  {journal} {\bibinfo  {journal} {Science}\
  }\textbf {\bibinfo {volume} {321}},\ \bibinfo {pages} {1306} (\bibinfo {year}
  {2008})}\BibitemShut {NoStop}%
\bibitem [{\citenamefont {Balents}(2010)}]{Balents2010}%
  \BibitemOpen
  \bibfield  {author} {\bibinfo {author} {\bibfnamefont {L.}~\bibnamefont
  {Balents}},\ }\href {\doibase 10.1038/nature08917} {\bibfield  {journal}
  {\bibinfo  {journal} {Nature}\ }\textbf {\bibinfo {volume} {464}},\ \bibinfo
  {pages} {199} (\bibinfo {year} {2010})}\BibitemShut {NoStop}%
\bibitem [{\citenamefont {Subramanian}\ \emph {et~al.}(1983)\citenamefont
  {Subramanian}, \citenamefont {Aravamudan},\ and\ \citenamefont {{G. V. Subba
  Rao}}}]{Subramanian1983}%
  \BibitemOpen
  \bibfield  {author} {\bibinfo {author} {\bibfnamefont {M.~A.}\ \bibnamefont
  {Subramanian}}, \bibinfo {author} {\bibfnamefont {G.}~\bibnamefont
  {Aravamudan}}, \ and\ \bibinfo {author} {\bibnamefont {{G. V. Subba Rao}}},\
  }\href {\doibase 10.1016/0079-6786(83)90001-8} {\bibfield  {journal}
  {\bibinfo  {journal} {Prog. Solid State Chem.}\ }\textbf {\bibinfo {volume}
  {15}},\ \bibinfo {pages} {55} (\bibinfo {year} {1983})}\BibitemShut {NoStop}%
\bibitem [{\citenamefont {Gardner}\ \emph {et~al.}(2010)\citenamefont
  {Gardner}, \citenamefont {Gingras},\ and\ \citenamefont
  {Greedan}}]{GardnerReview}%
  \BibitemOpen
  \bibfield  {author} {\bibinfo {author} {\bibfnamefont {J.~S.}\ \bibnamefont
  {Gardner}}, \bibinfo {author} {\bibfnamefont {M.~J.~P.}\ \bibnamefont
  {Gingras}}, \ and\ \bibinfo {author} {\bibfnamefont {J.~E.}\ \bibnamefont
  {Greedan}},\ }\href@noop {} {\bibfield  {journal} {\bibinfo  {journal} {Rev.
  Mod. Phys.}\ }\textbf {\bibinfo {volume} {82}},\ \bibinfo {pages} {53}
  (\bibinfo {year} {2010})}\BibitemShut {NoStop}%
\bibitem [{\citenamefont {Lee}\ \emph {et~al.}(2000)\citenamefont {Lee},
  \citenamefont {Broholm}, \citenamefont {Kim}, \citenamefont {Ratcliff},\ and\
  \citenamefont {Cheong}}]{Lee2000}%
  \BibitemOpen
  \bibfield  {author} {\bibinfo {author} {\bibfnamefont {S.-H.}\ \bibnamefont
  {Lee}}, \bibinfo {author} {\bibfnamefont {C.}~\bibnamefont {Broholm}},
  \bibinfo {author} {\bibfnamefont {T.~H.}\ \bibnamefont {Kim}}, \bibinfo
  {author} {\bibfnamefont {W.}~\bibnamefont {Ratcliff}}, \ and\ \bibinfo
  {author} {\bibfnamefont {S.-W.}\ \bibnamefont {Cheong}},\ }\href {\doibase
  10.1103/PhysRevLett.84.3718} {\bibfield  {journal} {\bibinfo  {journal}
  {Phys. Rev. Lett.}\ }\textbf {\bibinfo {volume} {84}},\ \bibinfo {pages}
  {3718} (\bibinfo {year} {2000})}\BibitemShut {NoStop}%
\bibitem [{\citenamefont {Chung}\ \emph {et~al.}(2005)\citenamefont {Chung},
  \citenamefont {Matsuda}, \citenamefont {Lee}, \citenamefont {Kakurai},
  \citenamefont {Ueda}, \citenamefont {Sato}, \citenamefont {Takagi},
  \citenamefont {Hong},\ and\ \citenamefont {Park}}]{Chung2005}%
  \BibitemOpen
  \bibfield  {author} {\bibinfo {author} {\bibfnamefont {J.-H.}\ \bibnamefont
  {Chung}}, \bibinfo {author} {\bibfnamefont {M.}~\bibnamefont {Matsuda}},
  \bibinfo {author} {\bibfnamefont {S.-H.}\ \bibnamefont {Lee}}, \bibinfo
  {author} {\bibfnamefont {K.}~\bibnamefont {Kakurai}}, \bibinfo {author}
  {\bibfnamefont {H.}~\bibnamefont {Ueda}}, \bibinfo {author} {\bibfnamefont
  {T.~J.}\ \bibnamefont {Sato}}, \bibinfo {author} {\bibfnamefont
  {H.}~\bibnamefont {Takagi}}, \bibinfo {author} {\bibfnamefont {K.-P.}\
  \bibnamefont {Hong}}, \ and\ \bibinfo {author} {\bibfnamefont
  {S.}~\bibnamefont {Park}},\ }\href {\doibase 10.1103/PhysRevLett.95.247204}
  {\bibfield  {journal} {\bibinfo  {journal} {Phys. Rev. Lett.}\ }\textbf
  {\bibinfo {volume} {95}},\ \bibinfo {pages} {247204} (\bibinfo {year}
  {2005})}\BibitemShut {NoStop}%
\bibitem [{\citenamefont {Harris}\ \emph {et~al.}(1997)\citenamefont {Harris},
  \citenamefont {Bramwell}, \citenamefont {McMorrow}, \citenamefont {Zeiske},\
  and\ \citenamefont {Godfrey}}]{Harris1997}%
  \BibitemOpen
  \bibfield  {author} {\bibinfo {author} {\bibfnamefont {M.~J.}\ \bibnamefont
  {Harris}}, \bibinfo {author} {\bibfnamefont {S.~T.}\ \bibnamefont
  {Bramwell}}, \bibinfo {author} {\bibfnamefont {D.~F.}\ \bibnamefont
  {McMorrow}}, \bibinfo {author} {\bibfnamefont {T.}~\bibnamefont {Zeiske}}, \
  and\ \bibinfo {author} {\bibfnamefont {K.~W.}\ \bibnamefont {Godfrey}},\
  }\href {\doibase 10.1103/PhysRevLett.79.2554} {\bibfield  {journal} {\bibinfo
   {journal} {Phys. Rev. Lett.}\ }\textbf {\bibinfo {volume} {79}},\ \bibinfo
  {pages} {2554} (\bibinfo {year} {1997})}\BibitemShut {NoStop}%
\bibitem [{\citenamefont {Ramirez}\ \emph {et~al.}(1999)\citenamefont
  {Ramirez}, \citenamefont {Hayashi}, \citenamefont {Cava}, \citenamefont
  {Siddharthan},\ and\ \citenamefont {Shastry}}]{Ramirez1999}%
  \BibitemOpen
  \bibfield  {author} {\bibinfo {author} {\bibfnamefont {A.~P.}\ \bibnamefont
  {Ramirez}}, \bibinfo {author} {\bibfnamefont {A.}~\bibnamefont {Hayashi}},
  \bibinfo {author} {\bibfnamefont {R.~J.}\ \bibnamefont {Cava}}, \bibinfo
  {author} {\bibfnamefont {R.}~\bibnamefont {Siddharthan}}, \ and\ \bibinfo
  {author} {\bibfnamefont {B.~S.}\ \bibnamefont {Shastry}},\ }\href@noop {}
  {\bibfield  {journal} {\bibinfo  {journal} {Nature}\ }\textbf {\bibinfo
  {volume} {399}},\ \bibinfo {pages} {333} (\bibinfo {year}
  {1999})}\BibitemShut {NoStop}%
\bibitem [{\citenamefont {Machida}\ \emph {et~al.}(2010)\citenamefont
  {Machida}, \citenamefont {Nakatsuji}, \citenamefont {Onoda}, \citenamefont
  {Tayama},\ and\ \citenamefont {Sakakibara}}]{Machida2010}%
  \BibitemOpen
  \bibfield  {author} {\bibinfo {author} {\bibfnamefont {Y.}~\bibnamefont
  {Machida}}, \bibinfo {author} {\bibfnamefont {S.}~\bibnamefont {Nakatsuji}},
  \bibinfo {author} {\bibfnamefont {S.}~\bibnamefont {Onoda}}, \bibinfo
  {author} {\bibfnamefont {T.}~\bibnamefont {Tayama}}, \ and\ \bibinfo {author}
  {\bibfnamefont {T.}~\bibnamefont {Sakakibara}},\ }\href@noop {} {\bibfield
  {journal} {\bibinfo  {journal} {Nature}\ }\textbf {\bibinfo {volume} {463}},\
  \bibinfo {pages} {210} (\bibinfo {year} {2010})}\BibitemShut {NoStop}%
\bibitem [{\citenamefont {Ross}\ \emph {et~al.}(2011)\citenamefont {Ross},
  \citenamefont {Savary}, \citenamefont {Gaulin},\ and\ \citenamefont
  {Balents}}]{Ross2011}%
  \BibitemOpen
  \bibfield  {author} {\bibinfo {author} {\bibfnamefont {K.~A.}\ \bibnamefont
  {Ross}}, \bibinfo {author} {\bibfnamefont {L.}~\bibnamefont {Savary}},
  \bibinfo {author} {\bibfnamefont {B.~D.}\ \bibnamefont {Gaulin}}, \ and\
  \bibinfo {author} {\bibfnamefont {L.}~\bibnamefont {Balents}},\ }\href
  {\doibase 10.1103/PhysRevX.1.021002} {\bibfield  {journal} {\bibinfo
  {journal} {Phys. Rev. X}\ }\textbf {\bibinfo {volume} {1}},\ \bibinfo {pages}
  {021002} (\bibinfo {year} {2011})}\BibitemShut {NoStop}%
\bibitem [{\citenamefont {Kimura}\ \emph {et~al.}(2013)\citenamefont {Kimura},
  \citenamefont {Nakatsuji}, \citenamefont {Wen}, \citenamefont {Broholm},
  \citenamefont {Stone}, \citenamefont {Nishibori},\ and\ \citenamefont
  {Sawa}}]{Kimura2013}%
  \BibitemOpen
  \bibfield  {author} {\bibinfo {author} {\bibfnamefont {K.}~\bibnamefont
  {Kimura}}, \bibinfo {author} {\bibfnamefont {S.}~\bibnamefont {Nakatsuji}},
  \bibinfo {author} {\bibfnamefont {J.-J.}\ \bibnamefont {Wen}}, \bibinfo
  {author} {\bibfnamefont {C.}~\bibnamefont {Broholm}}, \bibinfo {author}
  {\bibfnamefont {M.~B.}\ \bibnamefont {Stone}}, \bibinfo {author}
  {\bibfnamefont {E.}~\bibnamefont {Nishibori}}, \ and\ \bibinfo {author}
  {\bibfnamefont {H.}~\bibnamefont {Sawa}},\ }\href@noop {} {\bibfield
  {journal} {\bibinfo  {journal} {Nat. Commun.}\ }\textbf {\bibinfo {volume}
  {4}},\ \bibinfo {pages} {1934} (\bibinfo {year} {2013})}\BibitemShut
  {NoStop}%
\bibitem [{\citenamefont {Zheng}\ \emph {et~al.}(2005)\citenamefont {Zheng},
  \citenamefont {Kubozono}, \citenamefont {Nishiyama}, \citenamefont
  {Higemoto}, \citenamefont {Kawae}, \citenamefont {Koda},\ and\ \citenamefont
  {Xu}}]{Zheng2005}%
  \BibitemOpen
  \bibfield  {author} {\bibinfo {author} {\bibfnamefont {X.~G.}\ \bibnamefont
  {Zheng}}, \bibinfo {author} {\bibfnamefont {H.}~\bibnamefont {Kubozono}},
  \bibinfo {author} {\bibfnamefont {K.}~\bibnamefont {Nishiyama}}, \bibinfo
  {author} {\bibfnamefont {W.}~\bibnamefont {Higemoto}}, \bibinfo {author}
  {\bibfnamefont {T.}~\bibnamefont {Kawae}}, \bibinfo {author} {\bibfnamefont
  {A.}~\bibnamefont {Koda}}, \ and\ \bibinfo {author} {\bibfnamefont {C.~N.}\
  \bibnamefont {Xu}},\ }\href {\doibase 10.1103/PhysRevLett.95.057201}
  {\bibfield  {journal} {\bibinfo  {journal} {Phys. Rev. Lett.}\ }\textbf
  {\bibinfo {volume} {95}},\ \bibinfo {pages} {057201} (\bibinfo {year}
  {2005})}\BibitemShut {NoStop}%
\bibitem [{\citenamefont {Canals}\ and\ \citenamefont
  {Lacroix}(1998)}]{Canals1998}%
  \BibitemOpen
  \bibfield  {author} {\bibinfo {author} {\bibfnamefont {B.}~\bibnamefont
  {Canals}}\ and\ \bibinfo {author} {\bibfnamefont {C.}~\bibnamefont
  {Lacroix}},\ }\href@noop {} {\bibfield  {journal} {\bibinfo  {journal} {Phys.
  Rev. Lett.}\ }\textbf {\bibinfo {volume} {80}},\ \bibinfo {pages} {2933}
  (\bibinfo {year} {1998})}\BibitemShut {NoStop}%
\bibitem [{\citenamefont {Tsunetsugu}(2001{\natexlab{a}})}]{Tsunetsugu2001}%
  \BibitemOpen
  \bibfield  {author} {\bibinfo {author} {\bibfnamefont {H.}~\bibnamefont
  {Tsunetsugu}},\ }\href@noop {} {\bibfield  {journal} {\bibinfo  {journal} {J.
  Phys. Soc. Jpn.}\ }\textbf {\bibinfo {volume} {70}},\ \bibinfo {pages} {640}
  (\bibinfo {year} {2001}{\natexlab{a}})}\BibitemShut {NoStop}%
\bibitem [{\citenamefont {Tsunetsugu}(2001{\natexlab{b}})}]{Tsunetsugu2001b}%
  \BibitemOpen
  \bibfield  {author} {\bibinfo {author} {\bibfnamefont {H.}~\bibnamefont
  {Tsunetsugu}},\ }\href {\doibase 10.1103/PhysRevB.65.024415} {\bibfield
  {journal} {\bibinfo  {journal} {Phys. Rev. B}\ }\textbf {\bibinfo {volume}
  {65}},\ \bibinfo {pages} {024415} (\bibinfo {year}
  {2001}{\natexlab{b}})}\BibitemShut {NoStop}%
\bibitem [{\citenamefont {Berg}\ \emph {et~al.}(2003)\citenamefont {Berg},
  \citenamefont {Altman},\ and\ \citenamefont {Auerbach}}]{Berg2003}%
  \BibitemOpen
  \bibfield  {author} {\bibinfo {author} {\bibfnamefont {E.}~\bibnamefont
  {Berg}}, \bibinfo {author} {\bibfnamefont {E.}~\bibnamefont {Altman}}, \ and\
  \bibinfo {author} {\bibfnamefont {A.}~\bibnamefont {Auerbach}},\ }\href@noop
  {} {\bibfield  {journal} {\bibinfo  {journal} {Phys. Rev. Lett.}\ }\textbf
  {\bibinfo {volume} {90}},\ \bibinfo {pages} {147204} (\bibinfo {year}
  {2003})}\BibitemShut {NoStop}%
\bibitem [{\citenamefont {Moessner}\ \emph {et~al.}(2006)\citenamefont
  {Moessner}, \citenamefont {Sondhi},\ and\ \citenamefont
  {Goerbig}}]{Moessner2006}%
  \BibitemOpen
  \bibfield  {author} {\bibinfo {author} {\bibfnamefont {R.}~\bibnamefont
  {Moessner}}, \bibinfo {author} {\bibfnamefont {S.~L.}\ \bibnamefont
  {Sondhi}}, \ and\ \bibinfo {author} {\bibfnamefont {M.~O.}\ \bibnamefont
  {Goerbig}},\ }\href {\doibase 10.1103/PhysRevB.73.094430} {\bibfield
  {journal} {\bibinfo  {journal} {Phys. Rev. B}\ }\textbf {\bibinfo {volume}
  {73}},\ \bibinfo {pages} {094430} (\bibinfo {year} {2006})}\BibitemShut
  {NoStop}%
\bibitem [{\citenamefont {Kim}\ and\ \citenamefont {Han}(2008)}]{JH-Kim2008}%
  \BibitemOpen
  \bibfield  {author} {\bibinfo {author} {\bibfnamefont {J.~H.}\ \bibnamefont
  {Kim}}\ and\ \bibinfo {author} {\bibfnamefont {J.~H.}\ \bibnamefont {Han}},\
  }\href {\doibase 10.1103/PhysRevB.78.180410} {\bibfield  {journal} {\bibinfo
  {journal} {Phys. Rev. B}\ }\textbf {\bibinfo {volume} {78}},\ \bibinfo
  {pages} {180410} (\bibinfo {year} {2008})}\BibitemShut {NoStop}%
\bibitem [{\citenamefont {Burnell}\ \emph {et~al.}(2009)\citenamefont
  {Burnell}, \citenamefont {Chakravarty},\ and\ \citenamefont
  {Sondhi}}]{Burnell2009}%
  \BibitemOpen
  \bibfield  {author} {\bibinfo {author} {\bibfnamefont {F.~J.}\ \bibnamefont
  {Burnell}}, \bibinfo {author} {\bibfnamefont {S.}~\bibnamefont
  {Chakravarty}}, \ and\ \bibinfo {author} {\bibfnamefont {S.~L.}\ \bibnamefont
  {Sondhi}},\ }\href {\doibase 10.1103/PhysRevB.79.144432} {\bibfield
  {journal} {\bibinfo  {journal} {Phys. Rev. B}\ }\textbf {\bibinfo {volume}
  {79}},\ \bibinfo {pages} {144432} (\bibinfo {year} {2009})}\BibitemShut
  {NoStop}%
\bibitem [{\citenamefont {Okamoto}\ \emph {et~al.}(2013)\citenamefont
  {Okamoto}, \citenamefont {Nilsen}, \citenamefont {Attfield},\ and\
  \citenamefont {Hiroi}}]{Okamoto2013}%
  \BibitemOpen
  \bibfield  {author} {\bibinfo {author} {\bibfnamefont {Y.}~\bibnamefont
  {Okamoto}}, \bibinfo {author} {\bibfnamefont {G.~J.}\ \bibnamefont {Nilsen}},
  \bibinfo {author} {\bibfnamefont {J.~P.}\ \bibnamefont {Attfield}}, \ and\
  \bibinfo {author} {\bibfnamefont {Z.}~\bibnamefont {Hiroi}},\ }\href
  {\doibase 10.1103/PhysRevLett.110.097203} {\bibfield  {journal} {\bibinfo
  {journal} {Phys. Rev. Lett.}\ }\textbf {\bibinfo {volume} {110}},\ \bibinfo
  {pages} {097203} (\bibinfo {year} {2013})}\BibitemShut {NoStop}%
\bibitem [{\citenamefont {Scheikowski}\ and\ \citenamefont
  {M\"{u}ller-Buschbaum}(1993)}]{Scheikowski1993}%
  \BibitemOpen
  \bibfield  {author} {\bibinfo {author} {\bibfnamefont {M.}~\bibnamefont
  {Scheikowski}}\ and\ \bibinfo {author} {\bibfnamefont {H.}~\bibnamefont
  {M\"{u}ller-Buschbaum}},\ }\href@noop {} {\bibfield  {journal} {\bibinfo
  {journal} {Z. Anorg. Allg. Chem.}\ }\textbf {\bibinfo {volume} {619}},\
  \bibinfo {pages} {559} (\bibinfo {year} {1993})}\BibitemShut {NoStop}%
\bibitem [{\citenamefont {Rabbow}\ and\ \citenamefont
  {M\"{u}ller-Buschbaum}(1996)}]{Rabbow1996}%
  \BibitemOpen
  \bibfield  {author} {\bibinfo {author} {\bibfnamefont {C.}~\bibnamefont
  {Rabbow}}\ and\ \bibinfo {author} {\bibfnamefont {H.}~\bibnamefont
  {M\"{u}ller-Buschbaum}},\ }\href@noop {} {\bibfield  {journal} {\bibinfo
  {journal} {Z. Anorg. Allg. Chem.}\ }\textbf {\bibinfo {volume} {622}},\
  \bibinfo {pages} {100} (\bibinfo {year} {1996})}\BibitemShut {NoStop}%
\bibitem [{BYb({\natexlab{a}})}]{BYbZOimpurity2}%
  \BibitemOpen
  \href@noop {} {}\bibinfo {note} {An impurity phase of ZnO is characterized by
  the peaks at $2\theta$ = $31.8^\circ$ and $36.2^\circ$, but other phases
  cannot be identified due to the small intensity}\BibitemShut {NoStop}%
\bibitem [{\citenamefont {Hutchings}(1964)}]{Hutchings1964}%
  \BibitemOpen
  \bibfield  {author} {\bibinfo {author} {\bibfnamefont {M.~J.}\ \bibnamefont
  {Hutchings}},\ }\href@noop {} {\bibfield  {journal} {\bibinfo  {journal}
  {Solid State Phys.}\ }\textbf {\bibinfo {volume} {16}},\ \bibinfo {pages}
  {227} (\bibinfo {year} {1964})}\BibitemShut {NoStop}%
\bibitem [{\citenamefont {Stevens}(1952)}]{Stevens1952}%
  \BibitemOpen
  \bibfield  {author} {\bibinfo {author} {\bibfnamefont {K.~W.~H.}\
  \bibnamefont {Stevens}},\ }\href@noop {} {\bibfield  {journal} {\bibinfo
  {journal} {Proc. Phys. Soc. London Sect. A}\ }\textbf {\bibinfo {volume}
  {65}},\ \bibinfo {pages} {209} (\bibinfo {year} {1952})}\BibitemShut
  {NoStop}%
\bibitem [{\citenamefont {Lea}\ \emph {et~al.}(1962)\citenamefont {Lea},
  \citenamefont {Leask},\ and\ \citenamefont {Wolf}}]{Lea1962}%
  \BibitemOpen
  \bibfield  {author} {\bibinfo {author} {\bibfnamefont {K.~R.}\ \bibnamefont
  {Lea}}, \bibinfo {author} {\bibfnamefont {M.~J.~M.}\ \bibnamefont {Leask}}, \
  and\ \bibinfo {author} {\bibfnamefont {W.~P.}\ \bibnamefont {Wolf}},\
  }\href@noop {} {\bibfield  {journal} {\bibinfo  {journal} {J. Phys. Chem.
  Solids}\ }\textbf {\bibinfo {volume} {23}},\ \bibinfo {pages} {1381}
  (\bibinfo {year} {1962})}\BibitemShut {NoStop}%
\bibitem [{\citenamefont {{J. H. Van Vleck}}(1932)}]{VanVleck1932}%
  \BibitemOpen
  \bibfield  {author} {\bibinfo {author} {\bibnamefont {{J. H. Van Vleck}}},\
  }\href@noop {} {\emph {\bibinfo {title} {{The Theory of Electronic and
  Magnetic Susceptibilities}}}}\ (\bibinfo  {publisher} {Oxford Univ. Press},\
  \bibinfo {address} {London},\ \bibinfo {year} {1932})\BibitemShut {NoStop}%
\bibitem [{BYb({\natexlab{b}})}]{BYbZOmultipole2}%
  \BibitemOpen
  \href@noop {} {}\bibinfo {note} {The success of this model suggests no
  significantly large contribution from the high-order multipolar interactions
  between Yb$^{3+}$ ions which have been discussed for the case of the
  pyrochlore Yb$_2$Ti$_2$O$_7$ (see Refs. 29 and 30).}\BibitemShut {Stop}%
\bibitem [{\citenamefont {Thompson}\ \emph {et~al.}(2011)\citenamefont
  {Thompson}, \citenamefont {McClarty}, \citenamefont {Ronnow}, \citenamefont
  {Regnault}, \citenamefont {Sorge},\ and\ \citenamefont
  {Gingras}}]{Thompson2011}%
  \BibitemOpen
  \bibfield  {author} {\bibinfo {author} {\bibfnamefont {J.~D.}\ \bibnamefont
  {Thompson}}, \bibinfo {author} {\bibfnamefont {P.~A.}\ \bibnamefont
  {McClarty}}, \bibinfo {author} {\bibfnamefont {H.~M.}\ \bibnamefont
  {Ronnow}}, \bibinfo {author} {\bibfnamefont {L.~P.}\ \bibnamefont
  {Regnault}}, \bibinfo {author} {\bibfnamefont {A.}~\bibnamefont {Sorge}}, \
  and\ \bibinfo {author} {\bibfnamefont {M.~J.~P.}\ \bibnamefont {Gingras}},\
  }\href@noop {} {\bibfield  {journal} {\bibinfo  {journal} {Phys. Rev. Lett.}\
  }\textbf {\bibinfo {volume} {106}},\ \bibinfo {pages} {187202} (\bibinfo
  {year} {2011})}\BibitemShut {NoStop}%
\bibitem [{\citenamefont {Hayre}\ \emph {et~al.}(2013)\citenamefont {Hayre},
  \citenamefont {Ross}, \citenamefont {Applegate}, \citenamefont {Lin},
  \citenamefont {Singh}, \citenamefont {Gaulin},\ and\ \citenamefont
  {Gingras}}]{Hayre2013}%
  \BibitemOpen
  \bibfield  {author} {\bibinfo {author} {\bibfnamefont {N.~R.}\ \bibnamefont
  {Hayre}}, \bibinfo {author} {\bibfnamefont {K.~A.}\ \bibnamefont {Ross}},
  \bibinfo {author} {\bibfnamefont {R.}~\bibnamefont {Applegate}}, \bibinfo
  {author} {\bibfnamefont {T.}~\bibnamefont {Lin}}, \bibinfo {author}
  {\bibfnamefont {R.~R.~P.}\ \bibnamefont {Singh}}, \bibinfo {author}
  {\bibfnamefont {B.~D.}\ \bibnamefont {Gaulin}}, \ and\ \bibinfo {author}
  {\bibfnamefont {M.~J.~P.}\ \bibnamefont {Gingras}},\ }\href {\doibase
  10.1103/PhysRevB.87.184423} {\bibfield  {journal} {\bibinfo  {journal} {Phys.
  Rev. B}\ }\textbf {\bibinfo {volume} {87}},\ \bibinfo {pages} {184423}
  (\bibinfo {year} {2013})}\BibitemShut {NoStop}%
\bibitem [{\citenamefont {Curnoe}(2008)}]{Curnoe2008}%
  \BibitemOpen
  \bibfield  {author} {\bibinfo {author} {\bibfnamefont {S.~H.}\ \bibnamefont
  {Curnoe}},\ }\href {\doibase 10.1103/PhysRevB.78.094418} {\bibfield
  {journal} {\bibinfo  {journal} {Phys. Rev. B}\ }\textbf {\bibinfo {volume}
  {78}},\ \bibinfo {pages} {094418} (\bibinfo {year} {2008})}\BibitemShut
  {NoStop}%
\bibitem [{\citenamefont {McClarty}\ \emph {et~al.}(2009)\citenamefont
  {McClarty}, \citenamefont {Curnoe},\ and\ \citenamefont
  {Gingras}}]{McClarty2009}%
  \BibitemOpen
  \bibfield  {author} {\bibinfo {author} {\bibfnamefont {P.}~\bibnamefont
  {McClarty}}, \bibinfo {author} {\bibfnamefont {S.}~\bibnamefont {Curnoe}}, \
  and\ \bibinfo {author} {\bibfnamefont {M.}~\bibnamefont {Gingras}},\
  }\href@noop {} {\bibfield  {journal} {\bibinfo  {journal} {J.\ Phys.\ Conf.\
  Ser.}\ }\textbf {\bibinfo {volume} {145}},\ \bibinfo {pages} {012032}
  (\bibinfo {year} {2009})}\BibitemShut {NoStop}%
\bibitem [{\citenamefont {Yamashita}\ and\ \citenamefont
  {Ueda}(2000)}]{Yamashita2000}%
  \BibitemOpen
  \bibfield  {author} {\bibinfo {author} {\bibfnamefont {Y.}~\bibnamefont
  {Yamashita}}\ and\ \bibinfo {author} {\bibfnamefont {K.}~\bibnamefont
  {Ueda}},\ }\href {\doibase 10.1103/PhysRevLett.85.4960} {\bibfield  {journal}
  {\bibinfo  {journal} {Phys. Rev. Lett.}\ }\textbf {\bibinfo {volume} {85}},\
  \bibinfo {pages} {4960} (\bibinfo {year} {2000})}\BibitemShut {NoStop}%
\bibitem [{\citenamefont {Tchernyshyov}\ \emph {et~al.}(2002)\citenamefont
  {Tchernyshyov}, \citenamefont {Moessner},\ and\ \citenamefont
  {Sondhi}}]{Tchernyshyov2002}%
  \BibitemOpen
  \bibfield  {author} {\bibinfo {author} {\bibfnamefont {O.}~\bibnamefont
  {Tchernyshyov}}, \bibinfo {author} {\bibfnamefont {R.}~\bibnamefont
  {Moessner}}, \ and\ \bibinfo {author} {\bibfnamefont {S.~L.}\ \bibnamefont
  {Sondhi}},\ }\href {\doibase 10.1103/PhysRevLett.88.067203} {\bibfield
  {journal} {\bibinfo  {journal} {Phys. Rev. Lett.}\ }\textbf {\bibinfo
  {volume} {88}},\ \bibinfo {pages} {067203} (\bibinfo {year}
  {2002})}\BibitemShut {NoStop}%
\end{thebibliography}
\end{document}